\documentclass[aps,prb,twocolumn,showpacs,superscriptaddress,groupedaddress]{revtex4}  
\usepackage{graphicx}  
\usepackage{dcolumn}   
\usepackage{bm}        
\usepackage{amssymb}   
\usepackage[dvipsnames,usenames]{color}
\usepackage{color}
\usepackage{pdfpages}

\begin{document}


\newcommand{\etal}{{\sl et al.}}
\newcommand{\ie}{{\sl i.e.}}
\newcommand{\sto}{SrTiO$_3$} 
\newcommand{\lto}{LaTiO$_3$}
\newcommand{\lao}{LaAlO$_3$}
\newcommand{\lno}{LaNiO$_3$}
\newcommand{\nith}{Ni$^{3+}$}
\newcommand{\nitw}{Ni$^{2+}$}
\newcommand{\otw}{O$^{2-}$}
\newcommand{\alo}{AlO$ _2 $}
\newcommand{\tio}{TiO$ _2 $}
\newcommand{\eg}{$e_{g}$}
\newcommand{\tg}{$t_{2g}$}
\newcommand{\dzt}{$d_{z^2}$}
\newcommand{\dxtyt}{$d_{x^2-y^2}$}
\newcommand{\dxy}{$d_{xy}$}
\newcommand{\dxz}{$d_{xz}$}
\newcommand{\dyz}{$d_{yz}$}
\newcommand{\egp}{$e'_{g}$}
\newcommand{\ag}{$a_{1g}$}
\newcommand{\mub}{$\mu_{\rm B}$}
\newcommand{\ef}{$E_{\rm F}$}
\newcommand{\alao}{$a_{\rm LAO}$}
\newcommand{\asto}{$a_{\rm STO}$}
\newcommand{\nst}{$N_{\rm STO}$}
\newcommand{\lnnlam}{(LNO)$_N$/(LAO)$_M$}

\title{Confinement-driven transitions between topological and Mott phases in  (LaNiO$_3$)$_N$/(LaAlO$_3$)$_M$(111) superlattices}
\author{David Doennig}
\affiliation{Section Crystallography, Department of Earth and Environmental Sciences  and Center of Nanoscience,
University of Munich, Theresienstr. 41, 80333 Munich, Germany}
\author{Warren E. Pickett}
\affiliation{Department of Physics, University of California Davis, One Shields Avenue, Davis, California 95616, U.S.A.}
\author{Rossitza Pentcheva}
\email{pentcheva@lrz.uni-muenchen.de}
\affiliation{Section Crystallography, Department of Earth and Environmental Sciences and Center of Nanoscience,
University of Munich, Theresienstr. 41, 80333 Munich, Germany}
\date{\today}

\begin{abstract}
A set of broken symmetry two-dimensional ground states are predicted  in (111)-oriented (LaNiO$_3$)$_N$/(LaAlO$_3$)$_M$ ($N$/$M$) superlattices, based on density functional theory (DFT) calculations including a Hubbard $U$ term. An unanticipated Jahn-Teller distortion with $d_{z^2}$ orbital polarization and a FM  Mott insulating (and multiferroic) phase emerges in the double perovskite (1/1), that shows strong susceptibility to strain-controlled orbital engineering. The LaNiO$_3$ bilayer with graphene topology has a switchable  multiferroic (ferromagnetic (FM) and ferroelectric) insulating ground state with inequivalent Ni sites. Beyond $N=3$ the confined LaNiO$_3$ slab undergoes a metal-to-insulator transition through a half-semimetallic phase with conduction originating from the interfaces. Antiferromagnetic arrangements allow combining motifs of the bilayer and  single trigonal layer band structures in designed artificial mixed phases.
\end{abstract}

\maketitle

Rare earth nickelates $R$NiO$_3$ ($R$NO), with formal $d^7$ configuration, exhibit intriguing properties, e.g. a temperature-driven metal-to-insulator transition (MIT), related to the strongly distorted perovskite structure and the size of rare earth ion $R$ ~\cite{catalan,Balents}.  The origin of MIT is strongly debated:  instead of the Jahn-Teller (JT) distortion that one may expect of an $e_g^1$ ion, charge order~\cite{CO_mazin}, a site-selective Mott transition~\cite{siteselective} or  a prosaic order-disorder origin~\cite{wepCOprl} have been discussed. 

Recently, LaNiO$_3$ (LNO), the only $R$NO representative that remains metallic at all temperatures~\cite{Torrance1992}, has been in the spotlight of research, due to the proposal  that a cuprate-like behavior can be stabilized when confined in a superlattice (SL) with a band insulator, e.g. LaAlO$_3$ (LAO)~\cite{khaliulin}. However, despite intensive efforts the selective \dxtyt\ or \dzt\ orbital polarization as a function of strain could only partially be realized~\cite{wu2013}. Instead,  DFT studies on (001) SLs indicate that both $e_g$ states contribute to the Fermi surface~\cite{Hansmann2009,Han2011,BlancaRomero2011}. Nevertheless, these (001) SLs have proven to be a fruitful playground to explore low-dimensional phenomena such as a MIT due to confinement and Coulomb interaction~\cite{Boris,liuPRBR,Freeland2011,BlancaRomero2011,fiorentini,frano}.

The possibility of topologically nontrivial behavior is currently shifting the interest from the much studied (001) stacking of AO/BO$_2$ planes to the (111)-perovskite superlattices with a B/AO$_3$ sequence. Theoretical work has concentrated on the LNO bilayer sandwiched between LAO, where two triangular NiO$_6$ octahedron layers  form a buckled honeycomb lattice.  Model Hamiltonian studies together with DFT calculations~\cite{Yang2011,Ruegg2011,Ruegg2012,Ruegg2013} have shown topological phases with a set of four symmetric (around band center) bands, 
two flat and two crossing, forming a Dirac point (DP) at K, with quadratic band touching points at 
$\Gamma$. First experiments~\cite{Middey2012} report the growth of \lnnlam(111) superlattices on mixed-terminated LAO(111) surfaces with 
sheet resistance and activated transport, characteristic more of semiconductors than the predicted Dirac-point semimetals~\cite{Yang2011,Ruegg2011,Ruegg2012}, 
necessitating a thorough understanding of the evolution of properties with LNO and spacer thickness.

Based on material-specific DFT calculations, we uncover a rich set of electronic states in \lnnlam(111) SLs  with varying thickness $N$ and $M$. These range from  a JT orbitally-polarized phase in the 1/1 SL, unanticipated so far in nickelates and showing strong sensitivity to strain controlled orbital engineering, to a DP Fermi surface where a band gap opens up due to symmetry breaking in 2/4 and, finally, an insulator-to-metal transition with increasing LNO thickness. Moreover, antiferromagnetic arrangements allow design of band structures combining features of the monolayer and bilayer system. We discuss the mechanisms driving these symmetry breaking transitions.    

To explore the origin of this rich behavior we have performed DFT calculations, using the all-electron full-potential linearized augmented plane wave method as 
implemented in the WIEN2k code~\cite{wien2k}. For the exchange-correlation functional we used the generalized 
gradient approximation (GGA)~\cite{pbe96}. Static local electronic correlations were taken into account in 
the GGA+$U$ method.~\cite{anisimov93,ylvisaker2009} Previous studies on nickelate bulk and superlattices have used $U$ values between 3-8 eV~\cite{CO_mazin,giovanetti,prosandeev2012}.  Gou {\it et al.}~\cite{gou2011}  calculated a self-consistent  $U_{eff}=5.74$~eV, close to the value of 5.7~eV derived from fitting to XAS and XPS data. The main results in this study are obtained with $U=5$~eV, $J=0.7$~eV (Ni $3d$), $U=8$~eV (La $4f$), but a systematic analysis of the influence of  $U$ is presented below. 
The lateral lattice constant is fixed to \alao =3.79~\AA, corresponding to growth on a LAO(111) substrate used in the experiments mentioned above~\cite{Middey2012}. Octahedral tilts and distortions were fully considered when relaxing atomic positions, whether constrained to undistorted P321 symmetry (D$_{3d}$ point group) or fully released to P1 symmetry.
\begin{figure}[t!]
\includegraphics[angle=270,scale=0.57]{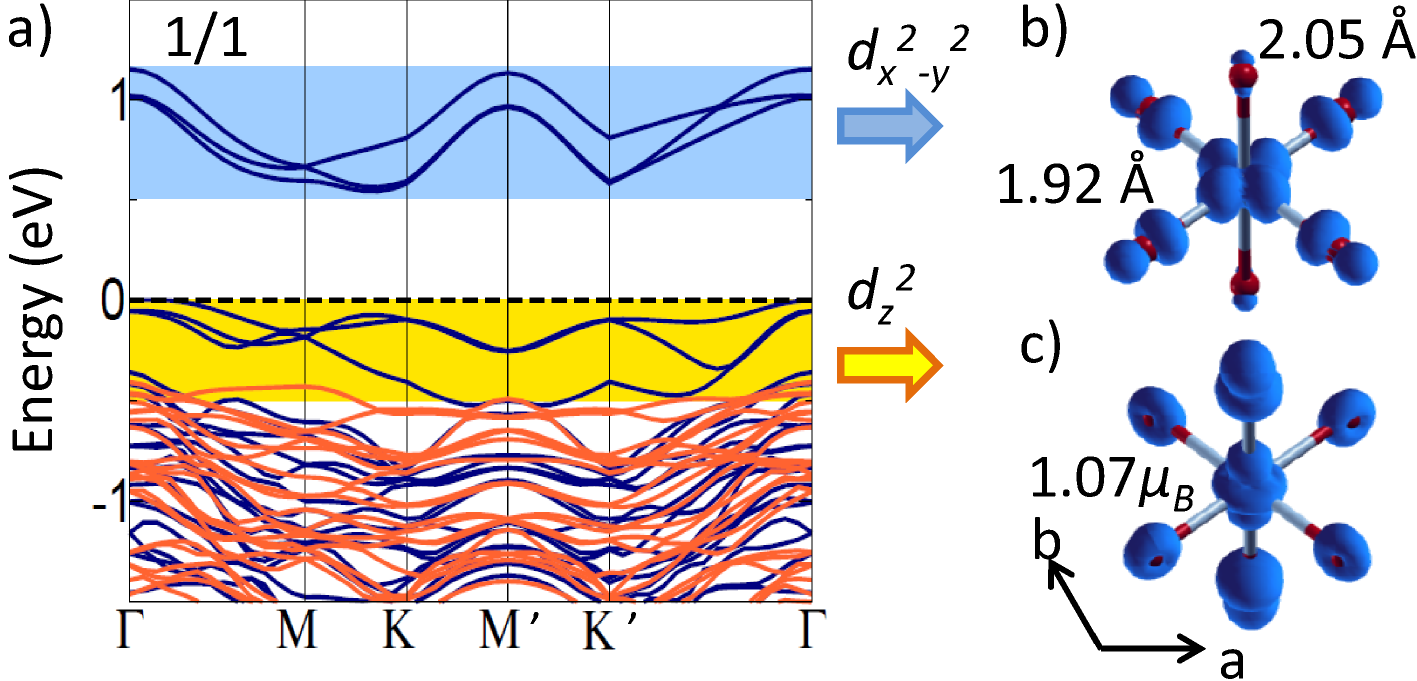}
\caption{\label{fig:1lno1lao} a) Band structure (red/blue correspond to minority/majority states)  and b-c) electron density distribution of the majority bands integrated in the shaded regions below  and above \ef\ in (1/1), demonstrating the $d_{z^2}$ (\dxtyt) character of the occupied (unoccupied) bands  and the ferro-orbital Jahn-Teller distorted insulator phase.}
\end{figure} 

The underlying symmetry, before charge, spin, or orbital ordering, is 
   C$_{3}$$\times$SU(2)$_{spin}$$\times$SU(2)$_{orbit}$$\times$Z$_2$,
 the latter expressing equivalence of the Ni sites (for $N>1$). We introduce the (heuristic) separation 
C$_3$$\times$Z$_2$$\leftrightarrow$D$_{3d}$ because
we will encounter these (broken) symmetries separately.
Consistent with earlier work, we find FM spin ordering to be favored in all cases studied (also with respect to non-collinear spin configurations), but we discuss some 
unusual metastable antiferromagnetic (AF) states later. As expected for Ni with an open $e_g$ shell, spin-orbit effects are found to be negligible~\cite{suppl}. 
We proceed to determine the superlattice behavior versus LNO thickness:

{\it 1/1}. An extreme and unanticipated orbital reconstruction occurs here, where the single Ni sheet forms a triangular 
lattice of second neighbors. Due to the ABC stacking of (111) planes in the fcc lattice, periodicity requires modelling in a (1/1)$_3$ unit cell, reflected in the set of three bands above and below \ef\ (Fig.~\ref{fig:1lno1lao}a).  The C$_{3}$ symmetry that protects $e_g$ orbital degeneracy is broken.  As Fig.~\ref{fig:1lno1lao}b-c show, an insulating FM ferro-orbital  phase emerges with $d_{z^2}$ orbital occupation  along one of the cubic (001)
axes, while the  $d_{x^2-y^2}$ orbital remains unoccupied. 
The orbital order (OO) is accompanied by a Jahn-Teller distortion of the octahedron, with 
axial (equatorial) Ni-O distance of 2.05 (1.92)~\AA. 
This finding is unexpected, given that no JT distortion has been reported in the $R$NiO$_3$ system~\cite{CO_mazin,siteselective}.  
 The mechanism of band gap opening in this FM OO Mott insulating (and multiferroic) state is distinct from the MIT in (001) oriented 
1LNO/1LAO SLs, where the insulating state emerges due to splitting into two inequivalent Ni sites under tensile strain~\cite{BlancaRomero2011,Freeland2011}.  

A more natural way to view this 1/1 system 
is as the double perovskite La$_2$NiAlO$_6$, where
the electropositive (non-transition metal, here Al) cation gives up its electrons and becomes a bystander to the fcc
sublattice of NiO$_6$ octahedra, as occurs in Ba$_2$NaOsO$_6$~\cite{BNOO}. 
The large $U$/$W$ ratio ($W$ is the bandwidth) enforces Mott physics, which is
complemented by the JT distortion with occupied $d_{z^2}$ symmetry 
`molecular orbitals' with large O participation and suppression of oxygen holes at the apical O (Fig.~\ref{fig:1lno1lao}b-c). 
The C$_3$ symmetry breaking due to the tetragonal JT distortion leads to inequivalent M, M' and K, K' points 
and to anisotropic $d_{z^2}$-$d_{z^2}$ hopping to neighbors. The bandwidths are correspondingly narrow and $k$-path dependent, 
0.15 eV along $\Gamma$-M and  0.5 eV along $\Gamma$-K.
\begin{figure}[t!]
\includegraphics[angle=0,scale=0.5]{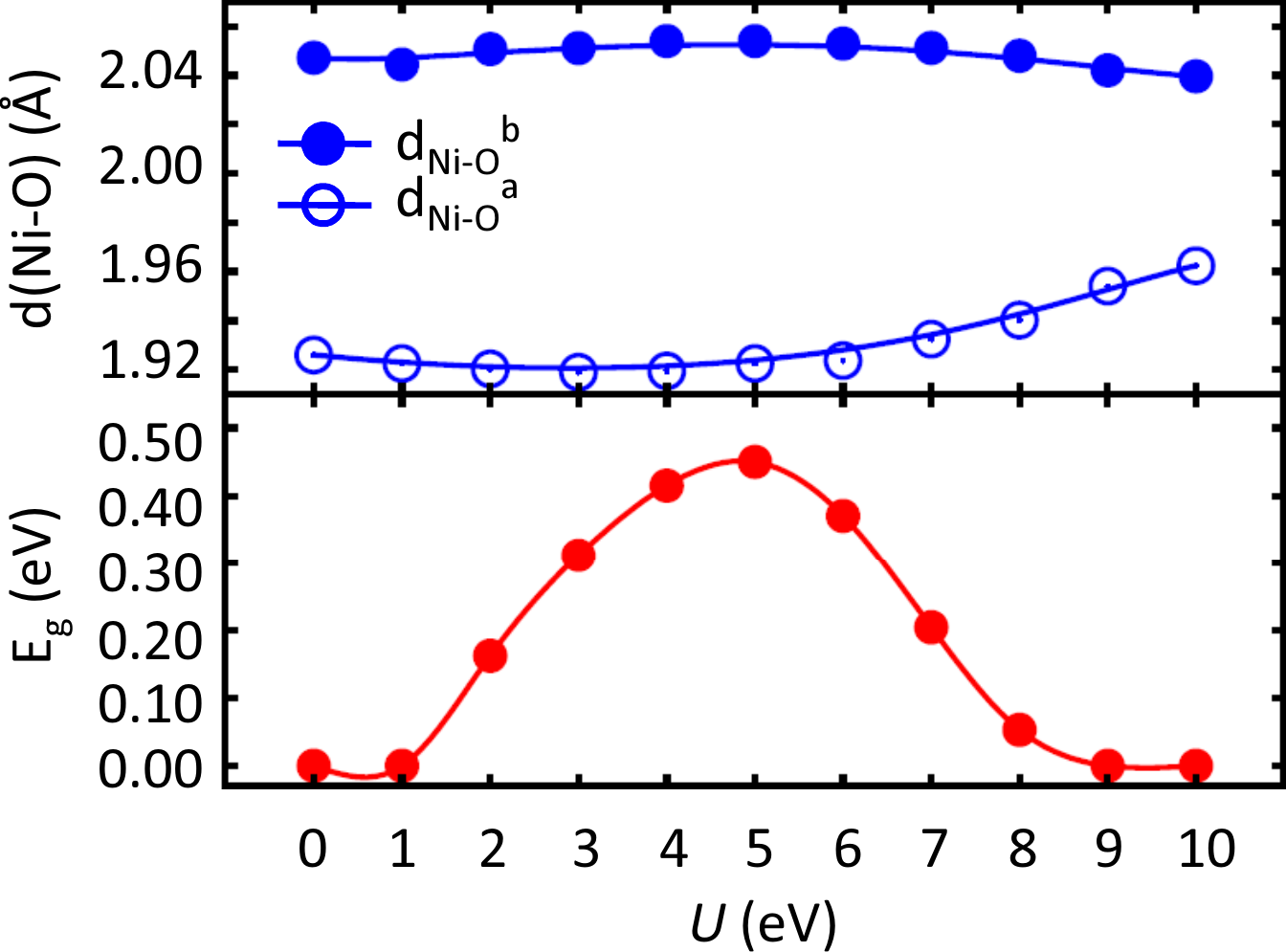}
\caption{\label{fig:U11} Ni-oxygen (apical and basal) distances (top) reflecting the Jahn-Teller distortion and band gap (bottom) as a function of $U$ for (LNO)$_1$/(LAO)$_1$(111). 
 }
\end{figure}
While the main features of the bands around \ef\ do not change dramatically, their separation and consequently the size of the band gap varies with $U$ (cf. Fig.~\ref{fig:U11}, bottom): for $U=0$~eV the system is metallic due to hole pockets at $\Gamma$. The band gap opens for $U=2$~eV and increases to  0.4~eV for $U=5$~eV, followed by a decrease and a band gap collapse at $U=9$~eV. This trend correlates with the size of Jahn-Teller distortion which also has a broad maximum around $U=5$~eV (cf. Fig.~\ref{fig:U11}, top).

Unlike the initial prediction of Chaloupka and Khaliullin~\cite{khaliulin}, only a weak orbital polarization was reported so far for (001)-oriented LNO  superlattices~\cite{BlancaRomero2011,Freeland2011,Han2011,Veenendaal2012}. Most recent experiments suggest exclusively positive values of $P = (n_{x^2-y^2}-n_{3z^2-r^2})/(n_{x^2-y^2} + n_{3z^2-r^2})$, e.g. +9$\%$ for SLs on SrTiO$_3$(001)~\cite{wu2013}. In contrast, our calculations demonstrate that strong control of orbital polarization can be achieved in strained La$_2$NiAlO$_6$ with switching between \dzt\ and \dxtyt\ orbital occupation for compressive (\alao) and tensile (\asto) strain within the (001) plane, respectively. Integrating over the antibonding states  between -3.0 and \ef, the orbital polarization values are 
$-44\%$ and +22$\%$, respectively. For comparison,  +9$\%$ were found for (001) SLs at \asto, using the same approach ~\cite{Han2011}. We attribute this enhanced affinity to strain-induced orbital engineering to the all Al nearest neighbors and the resulting partial suppression of Ni-O covalency in the double perovskite as opposed to the (001) SL.

\begin{figure}[t!]
\includegraphics[angle=270,scale=0.54]{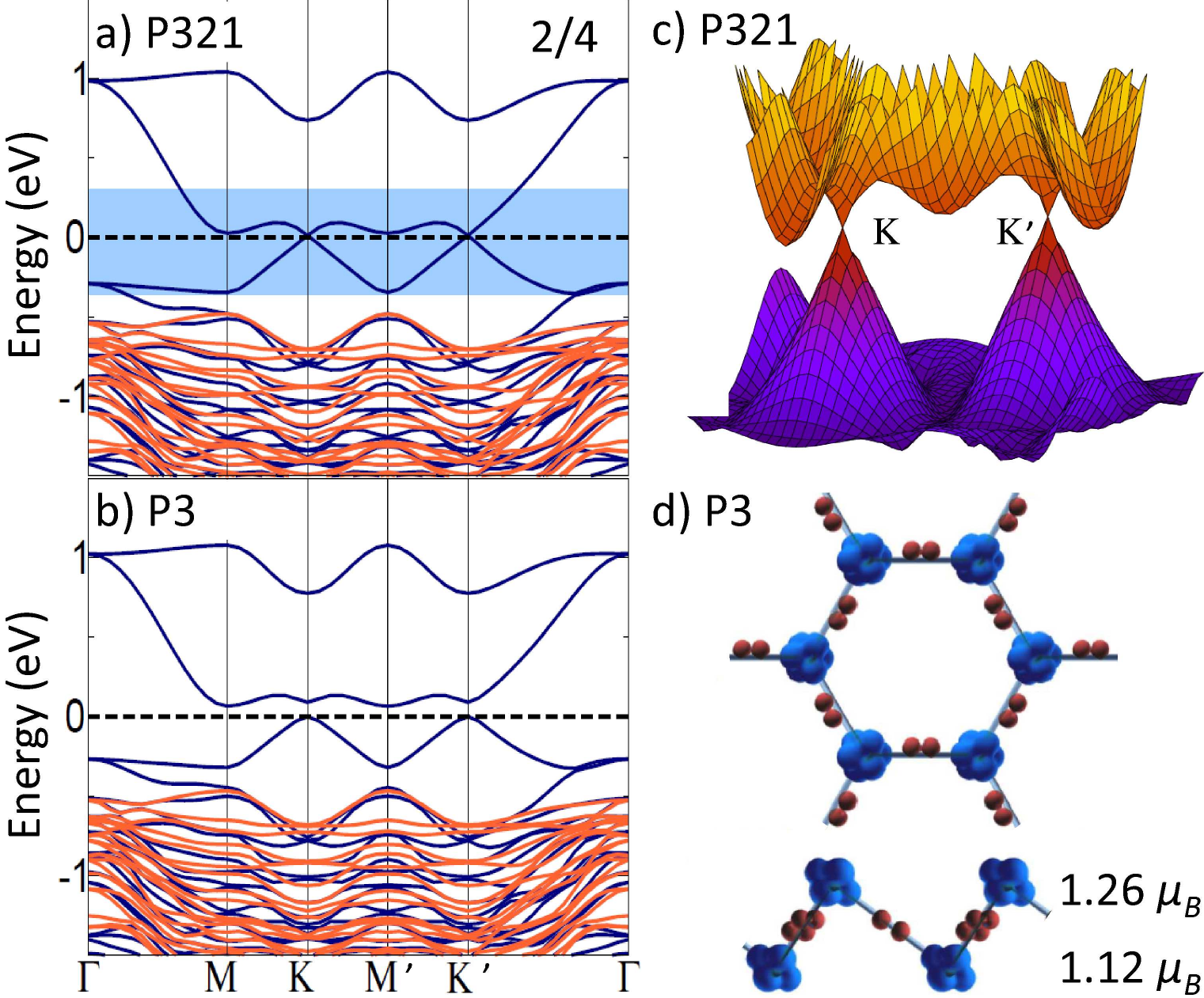}
\caption{\label{fig:2lno4lao} Majority (blue) and minority (red) band structures for the FM bilayer 2LNO/4LAO(111), within a) P321 and b) P3 symmetry. Symmetry breaking into two inequivalent Ni sites destroys the Dirac point by opening a gap of 0.06~eV at $K$ and $K'$; c) three dimensional band structure showing the Dirac cone around $K$ and $K'$; d) top and a side view of the
spin density distribution for P3.  
Note the disproportionation of Ni magnetic moments at the two interfaces and the strong contribution of O $2p$ states to the spin density.    
}
\end{figure}

{\it 2/4 Ni bilayer.}  The 4LAO slab isolates the Ni bilayer to a two dimensional system.
Constraining lattice symmetry to P321, we reproduce the previously reported FM Dirac-point half-semimetal~\cite{Ruegg2011,Yang2011}, shown in Fig.~\ref{fig:2lno4lao}a), that is characterized~\cite{Ruegg2012} by first and second neighbor hopping $t$=0.6~eV, $t'/t \approx 0.1$. Allowing full lattice relaxation  breaks `inversion' symmetry Z$_2$ and reduces the symmetry to P3, resulting in two inequivalent interfaces (Fig.~\ref{fig:2lno4lao}b): 
the Ni magnetic moment, 1.20~\mub\ within P321, becomes 1.12 and 1.26~\mub\ within  P3. The asymmetry at the two interfaces is also reflected in NiO$_6$ octahedra with mildly different average Ni-O bond lengths of 1.93 vs. 1.95~\AA, respectively. This primarily breathing distortion (reminiscent of the bulk $R$NO  and (LNO)$_1$/(LAO)$_1$(001) SLs~\cite{BlancaRomero2011}) opens a gap of 0.06~eV at the $K$ point. 

In the gapped state the system becomes multiferroic (ferroelectric and FM) with a great difference between majority and minority bandgaps. The effect is similar to, but much weaker than in the $t_{2g}$ system (SrTiO$_3$)$_2$/(LAO)$_4$(111) where `charge order' opens a gap of 0.7 eV~\cite{doennig}. The spin density, displayed in Fig.~\ref{fig:2lno4lao}d shows a mixed $d_{z^2}$ and  $d_{x^2-y^2}$ character (hence weak orbital polarization) at the Ni sites as observed previously also for (001) oriented LNO/LAO SLs~\cite{BlancaRomero2011,Han2011} and a hybridization with O $2p$ states.  Similar to the (001) oriented superlattices~\cite{Han2010,BlancaRomero2011,Freeland2011}, the O $2p$ hole density (not shown here) connecting to Al across the interface is suppressed. 

Calculations for a 2/2 superlattice display an overall similar band structure to the 2/4 system (Fig.~\ref{fig:2lno4lao}), indicating that coupling of the LNO layers through LAO is largely suppressed already for $M=2$. 
\begin{figure}[t!]
\includegraphics[angle=0,scale=0.5]{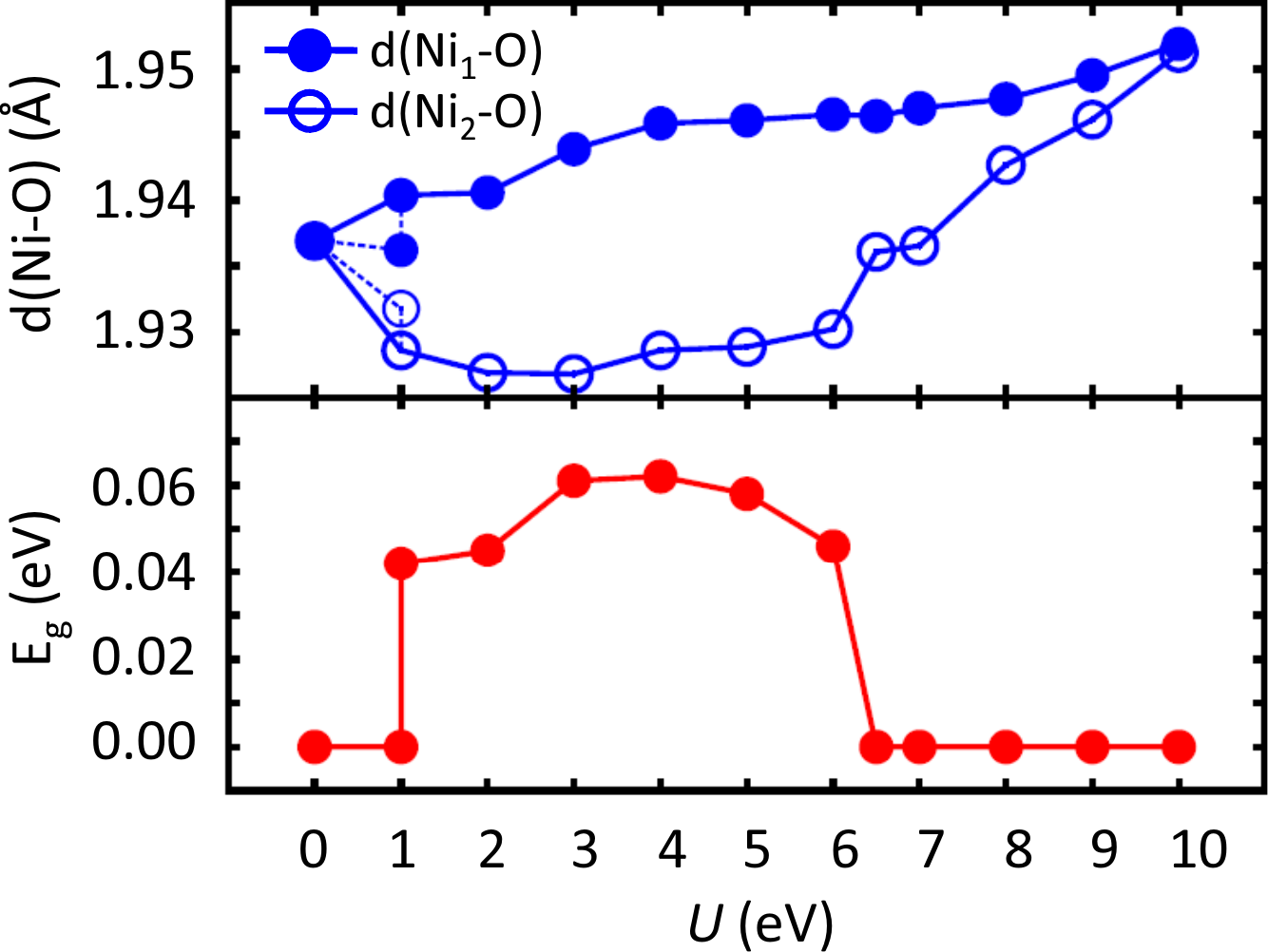}
\caption{\label{fig:U24} Breathing mode distortion of the two Ni sites (top) and band gap (bottom) as a function of $U$ for (LNO)$_2$/(LAO)$_4$(111). 
 }
\end{figure}

\begin{figure*}[t!]
\includegraphics[angle=270,scale=0.7]{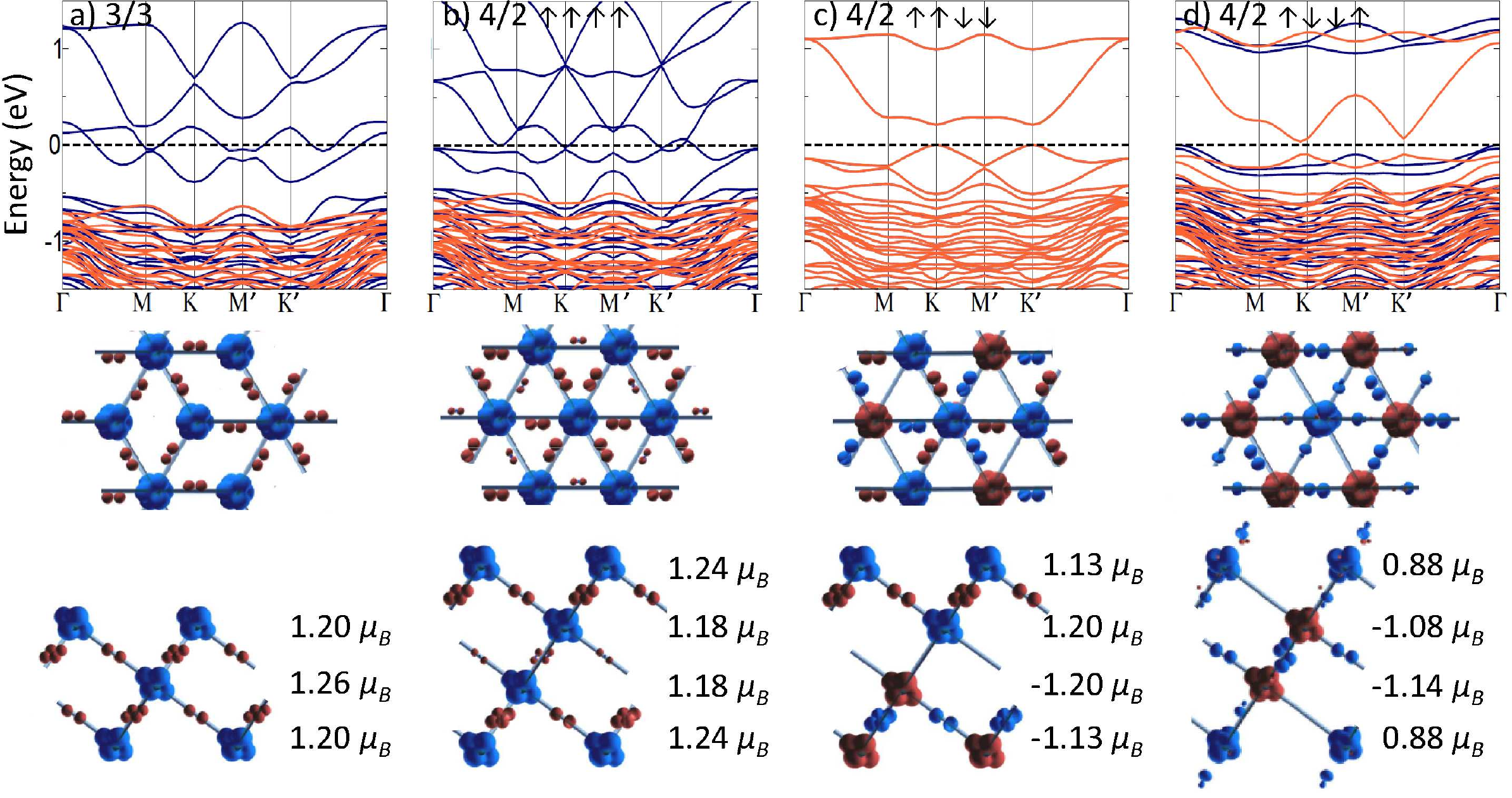}
\caption{\label{fig:nlnomlao} (Above) Majority (blue) and minority (red) band structures 
and (below) top and side view of the spin density distribution for $N=3-4$, in \lnnlam(111). 
a), b) crossover to a FM semimetallic state for $N=3,4$ with exclusively interface Ni contribution to the bands at \ef. c-d) AF coupling of the Ni layers for $N=4$  $\uparrow\uparrow\downarrow\downarrow$ and $\uparrow\downarrow\downarrow\uparrow$. Note that the former resembles the band structures of 2/4. (Fig.~\ref{fig:2lno4lao}b), while the latter combines the band structures of a bilayer (2/4) and a single triangular layer (1/1) with a gap between the flat red bands of 1.0~eV. Red/blue corresponds to minority/majority states.
 }
\end{figure*}
We find that the Dirac-point band structure and spin polarization are robust with respect to variation of $U$ in a broad window and persist even for $U=0$ eV, the FM state being slightly more stable than the non-magnetic system. The main effect of $U$ is the enhancement of exchange splitting and the narrowing of the relevant bands around \ef. In contrast, the Coulomb repulsion term is a driving force for opening the band gap due to symmetry breaking  and charge disproportionation of Ni in the bilayer structure: a first order transition takes place at $\sim 1$~eV to a gapped phase that persists up to $6-7$eV, beyond which the band gap is closed and the ``disproportionation'', as reflected in a breathing mode, is quenched (cf. Fig.~\ref{fig:U24}). Similarly, a suppression of disproportionation was recently reported for bulk NdNiO$_3$ above $U=8$ eV with a transition to a metallic spin-spiral state~\cite{prosandeev2012}.

{\it 3/3 and 4/2}. 
Increasing the LNO thickness to the dice lattice $N=3$ (Fig.~\ref{fig:nlnomlao}a) and to $N=4$ (Fig.~\ref{fig:nlnomlao}b) enhances the bandwidth and  leads to an  insulator-to-metal transition with a critical thickness $N_{c}=3$. 
(LNO)$_3$/(LAO)$_3$ has a slight overlap of topologically disjoint bands at \ef, so the MIT with thickness proceeds
through this half-semimetallic phase.
In both cases the orbital degeneracy of $e_g$ states (trigonal symmetry) is retained. 
Bands of the central Ni layers are  shifted from \ef, 
leaving fully polarized conduction to occur mainly through $e_g$ states of interface Ni, {\it i.e.} two barely separated parallel fully spin-polarized two-dimensional electron gases (2DEG).

{\it AF configurations.}  
Bulk nickelates $R$NiO$_3$ (except $R$=La) exhibit a magnetic ground state with a four layer repeat $\uparrow\uparrow\downarrow\downarrow$ along the  (111) direction where $\uparrow$/$\downarrow$ denotes magnetic moments  oriented up/down ~\cite{magnRNO}. Related to this bulk magnetic order, we have investigated two metastable AF configurations for the 4/2 superlattice, $\uparrow\uparrow\downarrow\downarrow$ and $\uparrow\downarrow\downarrow\uparrow$, 73 and 76~meV/Ni-Ni bond higher in energy than the FM 4/2, respectively. These energy differences involve lattice relaxations so they cannot be interpreted directly as magnetic
exchange energies. Further noncollinear arrangements are discussed in Suppl. Material~\cite{suppl}The bands reflect weak electronic coupling between neighboring $\uparrow$ and $\downarrow$ layers, due to the energy mismatch of bands of opposite spin directions. Interestingly, the band structure of $\uparrow\uparrow\downarrow\downarrow$ (Fig.~\ref{fig:nlnomlao}c) resembles that of  2/4,  rather than the FM 4/2 ($\uparrow\uparrow\uparrow\uparrow$), indicating that $\uparrow\uparrow\downarrow\downarrow$ can be considered as constructed from two oppositely oriented FM honeycomb bilayers $\uparrow\uparrow$ and $\downarrow\downarrow$ that are weakly coupled through the $\uparrow\downarrow$ link in the center. Asymmetry of the Ni sites in the outer/inner layers, reflected in the slight difference in magnetic moments (1.13 and 1.20~\mub), results in a gap at $K$, somewhat larger than in the case of 2/4 (and 2/2) and makes this system an AF Peierls insulator comprised of two spin-antialigned bilayers. 

Another curious case is $\uparrow\downarrow\downarrow\uparrow$ (Fig.~\ref{fig:nlnomlao}d), which is a low net moment ferrimagnet because $\uparrow$ and $\downarrow$ moments are inequivalent due to the layering. Here the band structure corresponds to an assemblage of the honeycomb bilayer ($\downarrow\downarrow$) weakly coupled to the two single triangular lattices ($\uparrow$) at the interfaces. The latter each produce a pair of occupied and unoccupied flat Hubbard bands of Mott insulating nature, separated by 1~eV.
This larger gap and its bandwidth indicates weaker interaction of the single $\uparrow$ layers through the antiparallel nickelate $\downarrow\downarrow$ bilayer than through the single LAO layer in 1/1 (Fig.~\ref{fig:1lno1lao}a). The $\downarrow\downarrow$ bilayer has the usual four band structure of 2/4, where the symmetry breaking between the two layers (magnetic moments of 1.08 and 1.14~\mub) opens a gap, but with one important difference: breaking of C$_{3}$ symmetry (in the single $\uparrow$ layers, similar to 1/1) results in the only case where  degeneracy of bands at $\Gamma$ is lifted. 

{\it Comments}.
The progression with LNO thickness thus proceeds as follows: With no Ni nearest neighbors, the 1/1 superlattice is a double
perovskite, with the narrow bandwidth driving a FM OO-JT Mott insulating phase with $d_{z^2}\uparrow$ orbitals
occupied and  all symmetries broken. This double perovskite is not only a rare realization of the JT effect in a nickelate system, but allows stronger control of orbital polarization with strain than so far achieved in the stoichiometrically equivalent (LNO)$_1$/(LAO)$_1$(001) superlattices. 
   
Whether separated by 2LAO or 4LAO, the (111) LNO bilayer is essentially the same 2D entity and becomes gapped
due to the breaking of the Z$_2$ symmetry of the two Ni sites with a band gap of 0.06~eV, in good agreement with the experimental value of  0.095~eV, obtained from transport data~\cite{Middey2012}. Since real charge order does not occur~\cite{wepCOprl}, this symmetry breaking transition is driven by 
electron-lattice coupling (Peierls mechanism), with some energy gain also from Hund's rule coupling.
 Our results on defect-free superlattices find Mott insulating phases of different origin for $N=1-2$, consistent with the high resistivities obtained in the transport measurements~\cite{Middey2012}, where some disorder is
expected for polar layer growth on a mixed termination substrate.  The 3LNO and 4LNO slabs are FM half-semimetal phases comprising a pair of fully spin-polarized 2DEG at the interfaces. Thus, with increasing thickness the system undergoes a MIT, approaching the bulk nickelate with structure clamped to the LAO (111) lattice constant. 

The rich spectrum of electronic phases uncovered in (111)-oriented LNO/LAO 
heterostructures as a function of LNO and LAO spacer thickness emerges due to symmetry breaking -- spin polarization, orbital ordering, and Ni site symmetry breaking
  -- and confinement.  
Atomic scale control of  the thickness of both constituents, strain  and gating-driven doping of the layers
opens possibilities to design artificial superlattices with exotic properties, from  (Mott
or Peierls) insulating, to parallel half-semimetallic 2DEGs, and possibly to topological phases.

\begin{acknowledgments}

R.P. and D. D. acknowledge financial support through the DFG SFB/TR80.
W. E. P. was supported by U.S. Department of Energy Grant No. DE-FG02-04ER46111.
\end{acknowledgments}

\end{document}